\documentclass[showpacs,aps,pre,twocolumn,superscriptaddress,textsuperscript]{revtex4}

\usepackage{graphicx}
\usepackage{amssymb}
\usepackage[sort&compress]{natbib}

\begin{document}

\title {Ring-like spin segregation of binary mixtures in a high-velocity rotating drum}

\author{Huang Decai}
\email[Corresponding author:] {hdc@mail.njust.edu.cn}
\affiliation{Department of Applied Physics, Nanjing University of Science and Technology, Nanjing 210094, China}

\author{Lu Ming}
\affiliation{School of Chemical Engineering, Nanjing University of Science and Technology,
Nanjing 210094, China}

\author{Sun Gang}
\email[Corresponding author:] {gsun@aphy.iphy.ac.cn}
\affiliation{Beijing National Laboratory for Condensed Matter Physics and
Key Laboratory of Soft Matter Physics, Institute of Physics, Chinese Academy of Sciences, Beijing 100190, China}

\author{Feng Yaodong}
\affiliation{Department of Applied Physics, Nanjing University of Science and Technology, Nanjing 210094, China}
\affiliation{Beijing National Laboratory for Condensed Matter Physics and
Key Laboratory of Soft Matter Physics, Institute of Physics, Chinese Academy of Sciences, Beijing 100190, China}

\author{Sun Min}
\affiliation{Department of Applied Physics, Nanjing University of Science and Technology,
Nanjing 210094, China}

\author{Wu Haiping}
\affiliation{Department of Applied Physics, Nanjing University of Science and Technology,
Nanjing 210094, China}

\author{Deng Kaiming}
\affiliation{Department of Applied Physics, Nanjing University of Science and Technology,
Nanjing 210094, China}

\date{\today}

\begin{abstract}
We present molecular dynamics simulations on the segregation of binary mixtures in a high-velocity rotating drum. Depending on the ratio of particle's radius and density, the ring-like spin segregation patterns in radial direction show similarities to the Brazil-nut effect and its reverse form. The smaller and heavier particles are easier to accumulate towards the drum wall and the bigger and lighter ones towards the drum center. Furthermore, we quantify the effects of particle's radius and density on the segregation states and plot the phase diagram of segregation in the ${\rho_b}/{\rho_s}$ - ${r_b}/{r_s}$ space. The observed phenomena can also be well explained by the combined actions, i.e., the percolation effect and the buoyancy effect.
\end{abstract}


\maketitle
\section{Introduction}

Granular materials are widespread in nature and daily life such as rocks, sands and sugars. The research on such a group of discrete particles is a particularly intriguing subject for engineers and physicists because of its gas, liquid and solid like properties \cite{Jaeger1996, Aranson2006}. A common example of segregation of granular mixtures usually happens ranging from industrial applications such as pharmaceutical processing and mineral transportation to natural phenomena such as snow avalanche and sand sedimentation in riverbeds \cite{Iverson1997, Bursik2005}. In the last decades, there have been a number of researches to control and characterize this peculiar property of granular mixtures. It is found that granular mixtures can segregate under a variety of conditions, i.e., shear \cite{Fan2011}, vibration \cite{Schnautz2005}, and rotation \cite{Hill2004}. The results show that the segregation states are governed by particle size \cite{Hong2001}, density \cite{Shi2007}, friction coefficient \cite{Ulrich2007}, vibration intensity \cite{Shi2009} and even the pressure gradient of air in the interstices between particles \cite{Yan2003}. Though large progresses have been made on this phenomenon, the behind physics remains still interesting and new behaviors or effects are to be found and investigated.

In general, the spontaneous segregation always happens during the granular flowing process. Thus the system of granule flowing in a horizontal rotating drum is a good candidate for the experimental and theoretical convenience \cite{Meier2008, Zuriguel2009}. Among these studies, two special cases of segregation of binary mixtures, where the particles differ in size (known as S-type system) and density (known as D-type system), have received intensive attentions \cite{Jain20051, Jain20052}. In an S-type system, the smaller particles are more easily sieved down though the interstices between particles, which is always called as the percolation effect. On the other hand, in a D-type system, the particles with higher density firstly sink down contrasted to the lighter ones, which is always named as the buoyancy effect. Furthermore, due to the combined actions of percolation and buoyancy effects in same direction, the degree of segregation can be reinforced in an S+D-type system, in which the bigger particles have a lower density and the smaller ones have a higher density. Reversely, the segregation state can be changed completely in an S-D-type system, in which the bigger particles have a higher density and the density driven effect takes an absolute role.

Many reported studies found that the rotational velocity is also a key control parameter on the segregation pattern \cite{Ottino2000, Seiden2007}. Depending on the rotational velocity of the drum, different flow regimes can be obtained, such as avalanching, rolling, cascading, cataracting, and centrifuging \cite{Rajchenbach1990, Meier2007}. Especially, Nityanand et al. found that the smaller particles accumulate in a classical core of the bed for the rolling regime \cite{Nityanand1986}. Such segregation can disappear when the granules flow in the cascading regime. Then the reversed segregation begins to occur when the granules flow in the cataracting regime. The segregation will become more pronounced under the centrifuging regime, in which the smaller granules adjacent to the drum wall are well graded and those having larger size far from the wall flow coarsely. To our knowledge, no much attentions associated with the centrifuging segregation of binary mixtures have been carried out. A more detailed study on the centrifuging granular flow in a rotational drum would be useful for a better understanding of the segregation of binary mixtures.

In this paper, we make a simulational study of the segregation of particles with different sizes and densities in a high-velocity rotating drum. In Sec. II, the simulation model is described and the segregation index is introduced to define the degree of segregation. In Sec. III, a series of systems, i.e., S-type, D-type, S+D-type and S-D-type systems, are simulated and the different ring-like spin segregation patterns are given. In Sec. IV, we quantify the effects of particle's radius and density on the segregation states and plot the phase diagram in the ${\rho_b}/{\rho_s}$ - ${r_b}/{r_s}$ space. Then a simple explanation consistent with the findings is discussed. Section V summarizes the conclusions of the work.

\section{Simulation model}

\begin{figure}[htbp]
\centering
\includegraphics[width=4.0cm, trim=0.0cm 0cm 0cm 0cm, clip]{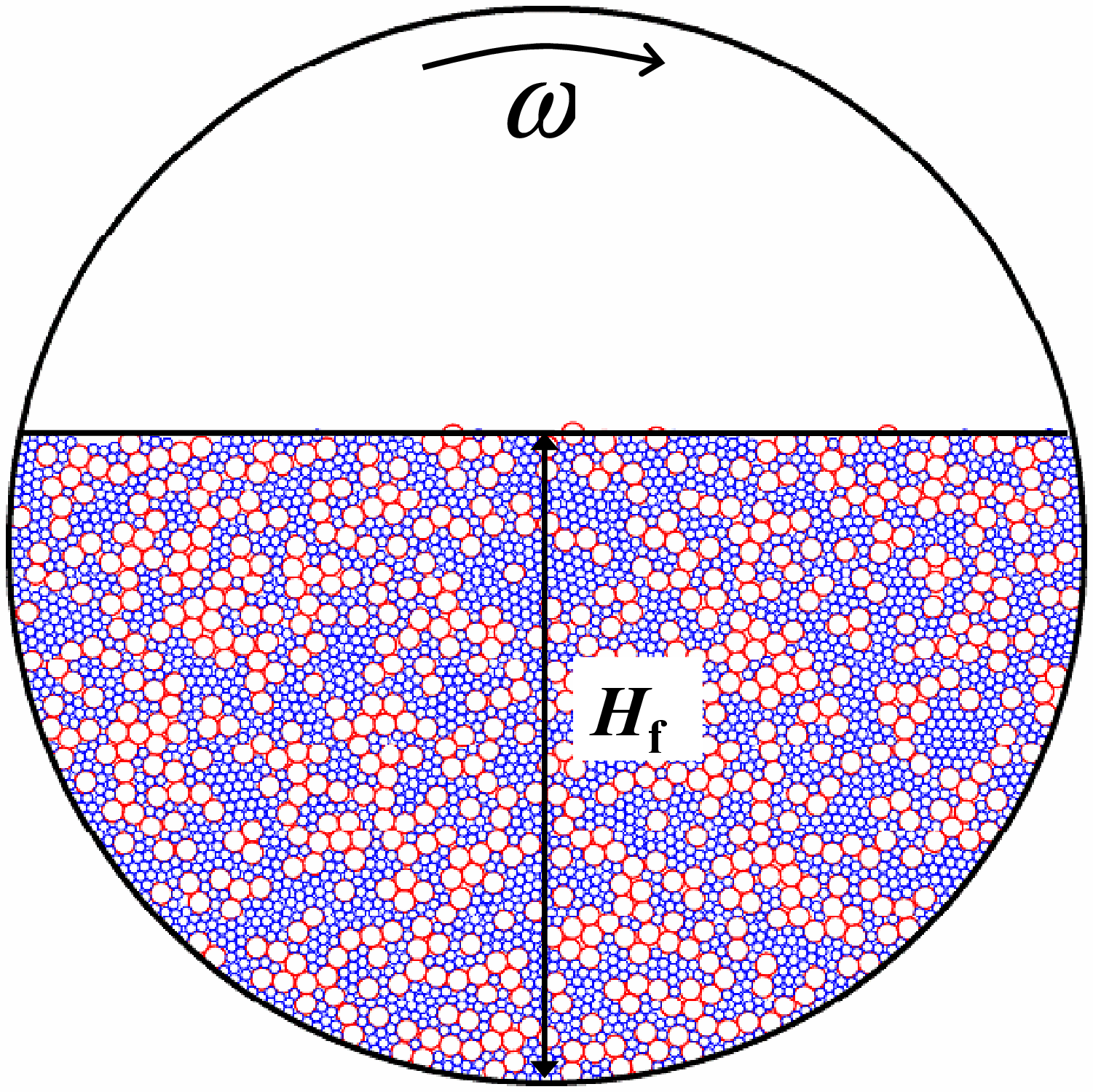}
\caption{A snapshot of system at the beginning of simulation.}
\label{fig:Figure1}
\end{figure}

The simulations are performed in a quasi two-dimensional drum as shown in Fig.\ref{fig:Figure1}. The drum is $D=200.0~{\rm mm}$ in diameter and $\delta_t=1.0~{\rm mm}$ in thickness. In this paper, all runs are begun from a randomly mixed state. The particles are randomly created according to their relative volumes in the upper part of drum and fall under the self-gravity when the rotational velocity is set to zero. The gravitational acceleration in the radial plane of drum is set to $g=9.8~{\rm m/s^2}$. Equal volumes of the two types of particles are used in all simulations. Depending on the rotational velocity, different flow regimes can be produced. Two indicated rotational frequency, $f_H$ and $f_L$, are critical for marking the stable flow bands of particles \cite{Juarez2011, Yang2008}. If the rotational frequency is much higher than $f_H$, all particles form a solid layer adhering to the drum wall due to the larger centrifugal force and finally rotate together with the drum. In this case, the particles have no relative motion and no clear segregation appears. If the rotational frequency is less than $f_L$, the majority of particles can not obtain enough energy to form a centrifuge flow. Here, $f_H$ and $f_L$ are values being about $100.0~{\rm rmp}$ and $200.0~{\rm rmp}$, which are also found to be closely correlated with the filling fraction and the friction coefficient between particles and between particles and drum wall. When the frequency is lowered to a moderate value between $f_H$ and $f_L$, some particles continue flowing along the drum wall, while other particles detach from the wall and flight as a free falling body. In the simulations, the filling height of particles is set to $H_f=0.6D$. The drum rotates clockwise around its center at a moderate high frequency, $f=150.0~{\rm rpm}$, corresponding to a angular velocity $\omega =15.71~{\rm s^{-1}}$.

Molecular dynamics simulation is adopted by using the discrete element method to describe the motion of each particle, which has been used in our previous studies \cite{Huang2006, Huang2011}. In a simulation time step, the positions and velocities of particles are updated by integrating the Newton's second law. To form a quasi two-dimensional system, the particles are circular in the radial direction of drum and the particle's thickness in the axial direction of drum is of the same value of drum thickness. In the simulation, both the translational motion in the radial plane of drum and the rotational motion in the axial direction of drum are considered. The force between two contacting particles is calculated in normal and tangential directions. The Kuwabara-Kono model is used to describe the normal interaction \cite{Kuwabara1987, Schafer1996},

\begin{equation}
F_{ij}^n=-k_n {\xi_{ij}}^{3/2} - \eta_n {\xi_{ij}}^{1/2} V_{ij}^n,
\label{eq:fn}
\end{equation}

The tangential component is determined by the minor of the tangential force with a memory effect and the force of dynamic friction,

\begin{equation}
F_{ij}^{\tau }=-\min(k_{t} {\zeta}_{ij}, \mu F_{ij}^n ),
\label{eq:ft}
\end{equation}

In Eqs.(\ref{eq:fn}) and (\ref{eq:ft}), $i$ and $j$ are the indexes of the particles. $\xi_{ij}=\max(0, d_{ij}-|{\bf x}_i-{\bf x}_j|)$ is the overlap of the contacting particles $i$ and $j$, $d$ is the sum of radiuses of the contacting particles $r_i$ and $r_j$. $\zeta_{ij}$ denotes the displacement in the tangential direction that took place since time $t_0$, when the contact was first established, i.e., $\zeta_{ij} (t)=\int_{t_0}^t v_s(t')dt'$. ${\bf V}_{ij}=V_{ij}^n {\bf e}_n + V_{ij}^{\tau} {\bf e}_{\tau}$ is the relative velocities of two particles. The indexes $n$ and $\tau$ represent the normal and tangential directions at the contact point, respectively. $k_n$, $k_t$ and $\eta_n $ characterize the stiffness and damping of the granular materials and are related to the collision time $t_n$ and the masses of contacting particles $m_i$ and $m_j$. The detailed values can be calculated as follows, $k_n=\frac {4}{3} \frac {{Y_i}{Y_j}}{Y_i+Y_j}\sqrt{\frac {{R_i}{R_j}}{R_i+R_j}}$, $k_t=\frac {2}{7} \frac {{m_i}{m_j}}{m_i+m_j}({\frac {\pi}{t_n}})^2$, $Y=\frac {E}{1-{\nu}^2}$, $t_n=3.21 ({\frac {m_i m_j}{m_i+m_j}})^{\frac {2}{5}} (V_{ij}^{n})^{- \frac {1}{5}}$. $E$ is the Young modulus, $\nu$ the Poisson ratio. The coefficient of sliding friction is set to $\mu =0.5$. The collisions between the particles and the wall of drum are treated as particle-particle collisions, except that the drum has an infinite mass and a diameter $D$.

In the simulation, the center position of the masses is easily calculated, $h_p={\frac {1}{N_p}} \sum\nolimits_{i=1}^{N_p} z_i $. The angular velocity of the masses is $\omega_p={\frac {1}{N_p}} \sum\nolimits_{i=1}^{N_p} \omega_i $. Here, $p$ is the species of particles and $N_p$ is the total number of $p$ species of particles. $z_i$ is the distance of particle $i$ with respect to the edge of drum. $\omega_i$ is the angular velocity of particle $i$. In order to have an quantified description of the degree of segregation, the segregation index $\chi$ is usually introduced \cite{Ciamarra2006},

\begin{equation}
\chi=2.0 \frac {h_b-h_s}{ h_b+h_s}
\label{eq:hbs}
\end{equation}

In Eq. (\ref {eq:hbs}), $h_b$ and $h_s$ are the center positions of the bigger particles and the smaller particles, respectively. As a comparison with the segregation in a vertically vibrated container \cite{Schnautz2005,Hong2001}, we also call the segregation state as the Brazil-nut effect (BNE), in which the bigger particles migrate towards the center of drum and the smaller ones towards the drum wall. If the segregated state is reversed, i.e., the bigger particles accumulate towards the drum wall and the smaller ones towards the center of drum, the corresponding segregated state is so named the Reverse Brazil-nut effect (RBNE). Apparently, the BNE happens for a positive $\chi$, while the RBNE for a negative $\chi$ according to the Eq. (\ref {eq:hbs}).

\section{Ring-like spin segregation pattern}

\begin{figure}[htbp]
\centering
\includegraphics[width=7.5cm]{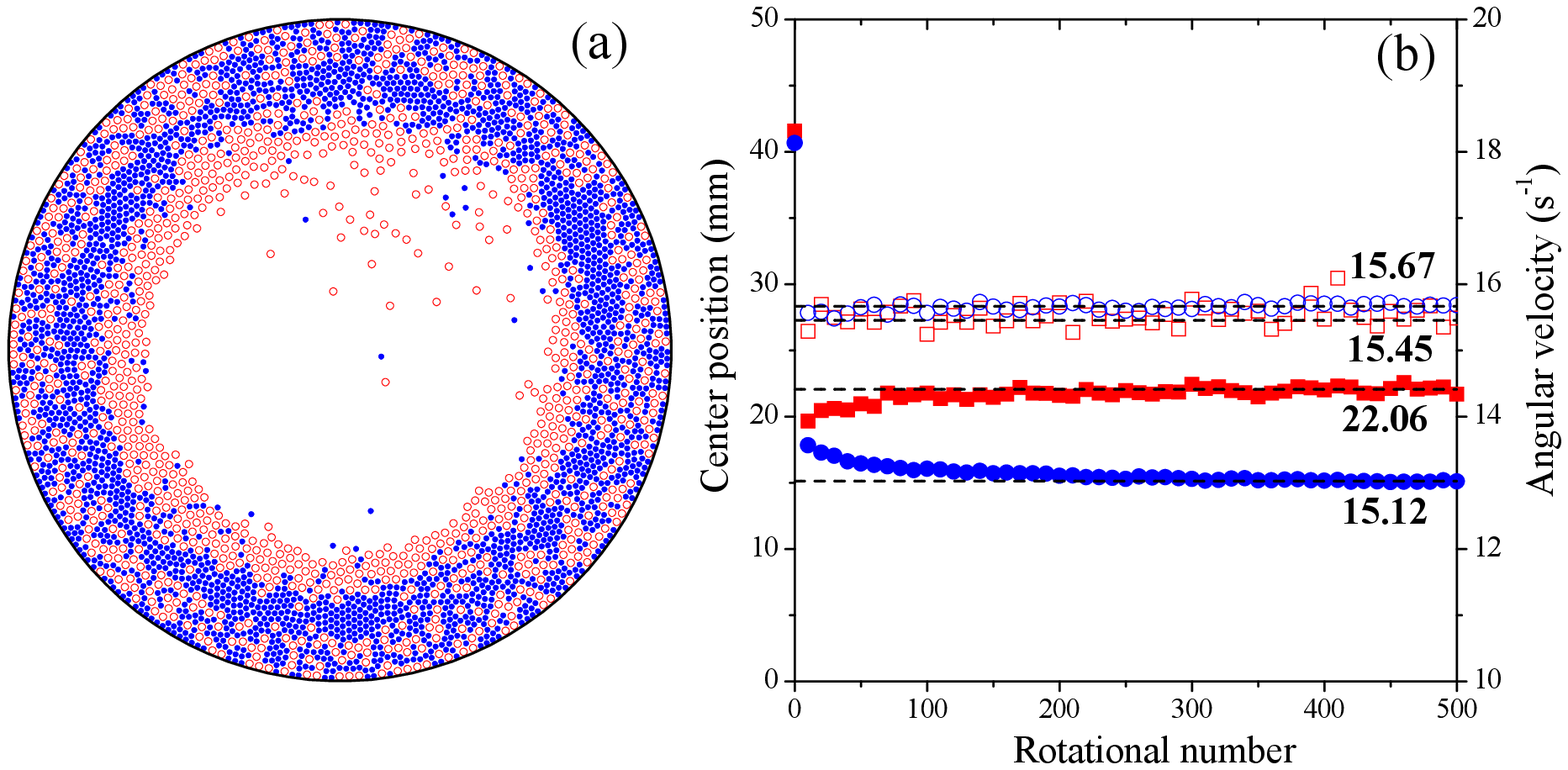}
\caption{(color online) (a) Image of the S-type system after 500 rotations. The densities of both bigger particles (open symbols) and smaller particles (solid symbols) are of the same value, $\rho_b=\rho_s=1.0 \times 10^3~{\rm kg/m^3}$. The radiuses of particles are for different sizes being $r_b=1.5~{\rm mm}$ and $r_s=1.0~{\rm mm}$, respectively. (b) Evolutions of the center positions (solid symbols) and the angular velocity (open symbols) of bigger particles and smaller particles corresponding to the system of (a). The square symbols and circle symbols are results for the bigger particles and smaller particles, respectively.}
\label{fig:Figure2}
\end{figure}

The simulations are firstly performed for an S-type system where the particles are of the same density but differ in size. Fig.\ref {fig:Figure2}(a) shows the segregation pattern of the binary mixtures after 500 rotations for $r_b=1.5~{\rm mm}$ bigger particles (red circles) and $r_s=1.0~{\rm mm}$ smaller particles (blue circles). We can see that two ring-like spin segregated patterns in the radial direction appear except a mixed region near the drum wall. The bigger particles migrate towards to the center of the drum and form a ring-like layer with a smaller radius, while the smaller particles surround at the periphery. In addition, the inner ring with bigger particles is actually a semi-circle, in which a few of bigger particles leave from the flowing layer and flight freely at the upper right region.

To give more insight on the segregation process of different species of particles, we plot the evolutions of the center position and the angular velocity of bigger particles and smaller particles as shown in Fig.\ref {fig:Figure2}(b). Fig.\ref {fig:Figure2}(b) shows that the segregation happens in the first hundred of rotations, i.e., the bigger particles are pushed towards the center of drum ($h_b=22.06~\rm mm$) and the smaller particles towards the drum wall ($h_s=15.12~\rm mm$). At the meantime, the averaged angular velocity of both the bigger particles and the smaller particles increases very fast and approaches nearly to that of the drum, $\omega_b=15.45~\rm s^{-1}$ and $\omega_s=15.67~\rm s^{-1}$. Further calculation shows that the particles at the mixed region just adjacent to the drum wall have the same angular velocity as that of the drum. The particles have no relative motions and this results in that the particles in the mixed region will not segregate even if the rotation continues.

\begin{figure}[htbp]
\centering
\includegraphics[width=7.5cm]{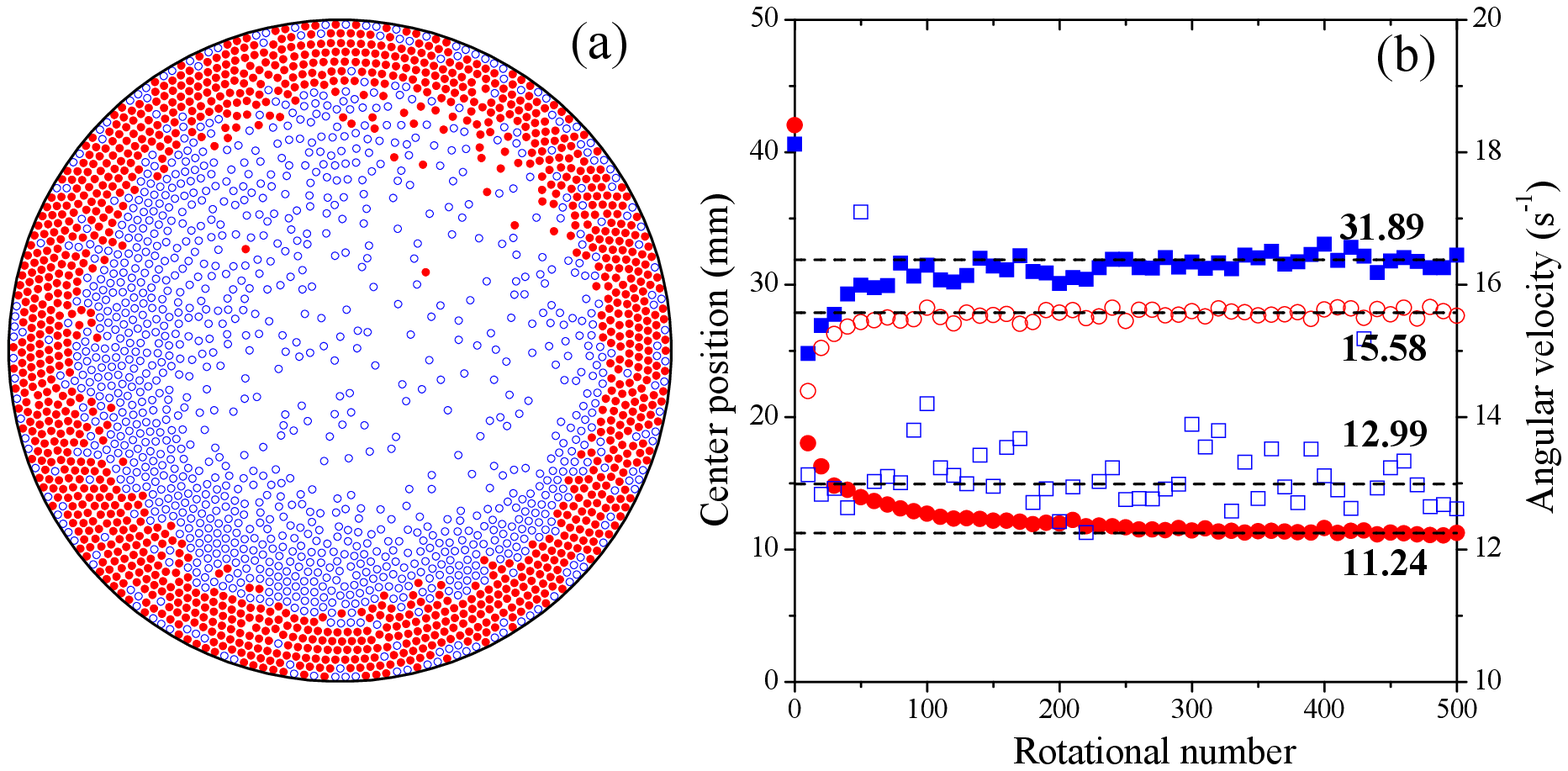}
\caption{(color online) (a) Image of the D-type system after 500 rotations. The radiuses of both heavier particles (solid symbols) and lighter particles (open symbols) are of the same size, $r_h=r_l=1.5~{\rm mm}$. The densities of particles are for different values being $\rho_h=3.0 \times 10^3~{\rm kg/m^3}$ and $\rho_l=1.0 \times 10^3~{\rm kg/m^3}$, respectively. (b) Evolutions of the center positions (solid symbols) and the angular velocity (open symbols) of bigger particles and smaller particles corresponding to the system of (a). The circle symbols and square symbols are results for heavier (lighter) particles.}
\label{fig:Figure3}
\end{figure}

Second, the simulations are carried out for a D-type system in which the particles are of the same size ($r_h= r_l=1.5~{\rm mm}$), but the densities of heavy particles and light particles are set to $\rho_h=3.0 \times 10^3~{\rm kg/m^{3}}$ and $\rho_l=1.0 \times 10^3~{\rm kg/m^{3}}$, respectively. The particles also segregate into two ring-like spin layers by density as shown in Fig.\ref {fig:Figure3}(a). The two rings have high concentrations with heavier particles locating at the periphery and lighter ones at the inner side. Likewise, the inner ring is a semi-circle, where some lighter particles detach from the flowing layer on the upper right region. As shown in Fig.\ref {fig:Figure3}(b), the segregation also proceeds quickly. The heavier particles migrate towards the drum wall ($h_h=11.24~\rm mm$) while the lighter ones towards the drum center ($h_l=31.89~\rm mm$). Furthermore, the heavier particles approaching closer to the drum wall have a higher angular velocity ($\omega_h=15.58~\rm s^{-1}$) compared to the lighter particles ($\omega_l=12.99~\rm s^{-1}$).

\begin{figure}[htbp]
\centering
\includegraphics[width=7.5cm]{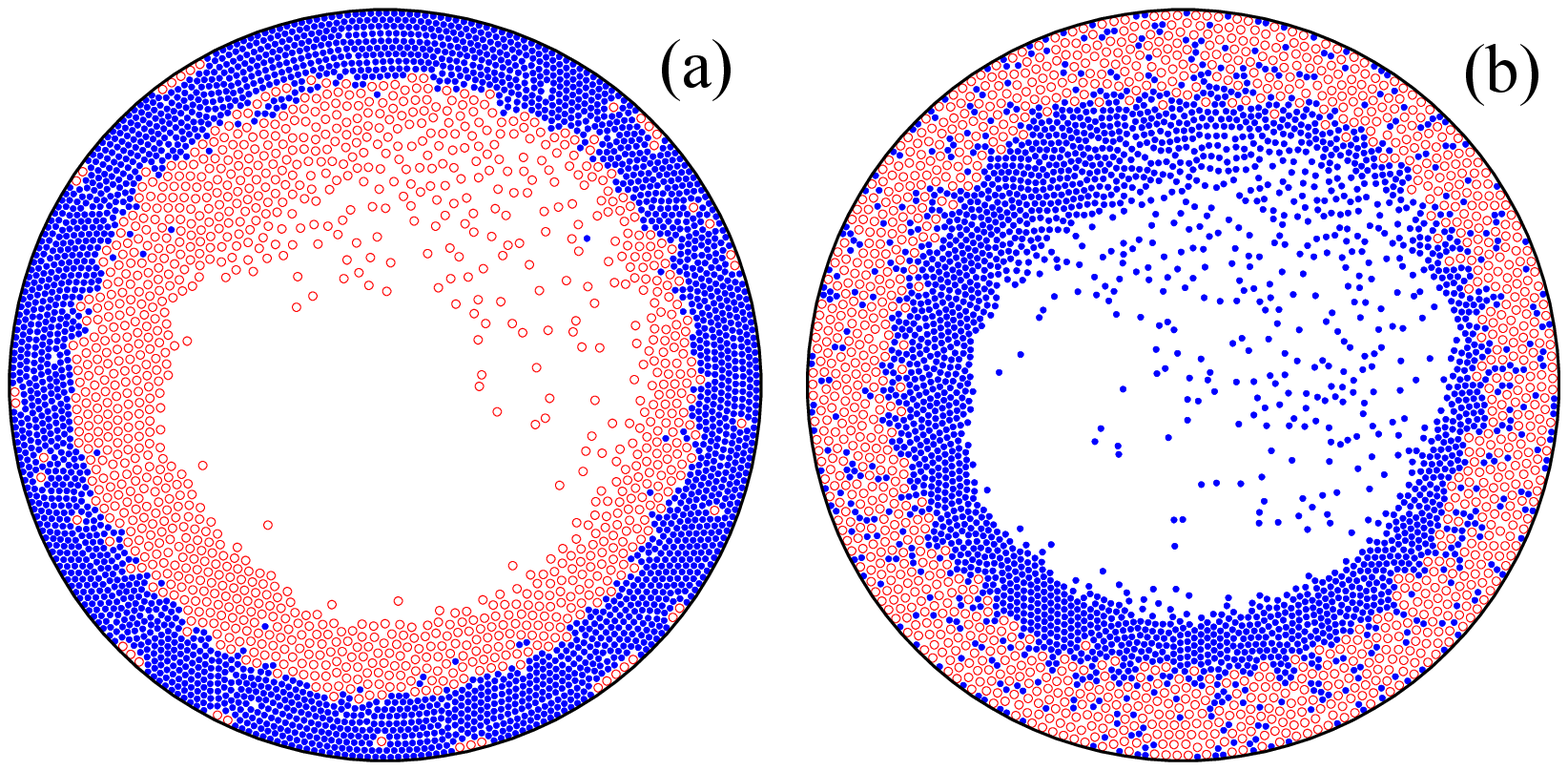}
\caption{(color online) Combined size and density segregation patterns of binary mixtures after 500 rotations. The radiuses of bigger particles (open symbols) and smaller ones (solid symbols) are for different sizes being $r_b=1.5~{\rm mm}$ and $r_s=1.0~{\rm mm}$, respectively. (a) An S+D-type system. The densities of bigger and smaller particles are set to $\rho_b=1.0 \times 10^3~{\rm kg/m^3}$ and $\rho_s=3.0 \times 10^3~{\rm kg/m^3}$, respectively. (b) An S-D-type system. The densities of bigger and smaller particles are set to $\rho_b=3.0 \times 10^3~{\rm kg/m^3}$ and $\rho_s=1.0 \times 10^3~{\rm kg/m^3}$, respectively.}
\label{fig:Figure4}
\end{figure}

A more complicated segregation happens for the cases where the binary mixtures are different in both size and density. We set fixedly the radiuses of bigger particles $r_b=1.5 ~{\rm mm}$ and smaller particles $r_s=1.0~{\rm mm}$, and then change the density ratio of bigger particles to smaller ones. Fig.\ref {fig:Figure4}(a) shows the segregation pattern after 500 rotations for an S+D-type system, where the densities of bigger particles and smaller particles are the values being $\rho_b=1.0 \times 10^3~{\rm kg/m^3}$ and $\rho_s=3.0 \times 10^3~{\rm kg/m^3}$, respectively. Compared to the segregation pattern in Fig.\ref {fig:Figure2}(a), we can see that two more clearer ring-like spin segregation layers appear, in which the smaller and heavier particles gather at the periphery and the bigger and lighter particles at the inner side. If the density of bigger particles is increased larger than that of smaller ones to form an S-D-type system, the segregation can be reversed completely, i.e., the bigger particles appear at the outer side and the smaller ones at the inner side. Fig.\ref {fig:Figure4}(b) shows a typical RBNE segregation pattern, where the density of bigger particles is set to $\rho_b=3.0 \times 10^3~{\rm kg/m^3}$, and that of smaller ones $\rho_s=1.0 \times 10^3~{\rm kg/m^3}$.

\section{Discussions}

\begin{figure}[htbp]
\centering
\includegraphics[width=7.0cm]{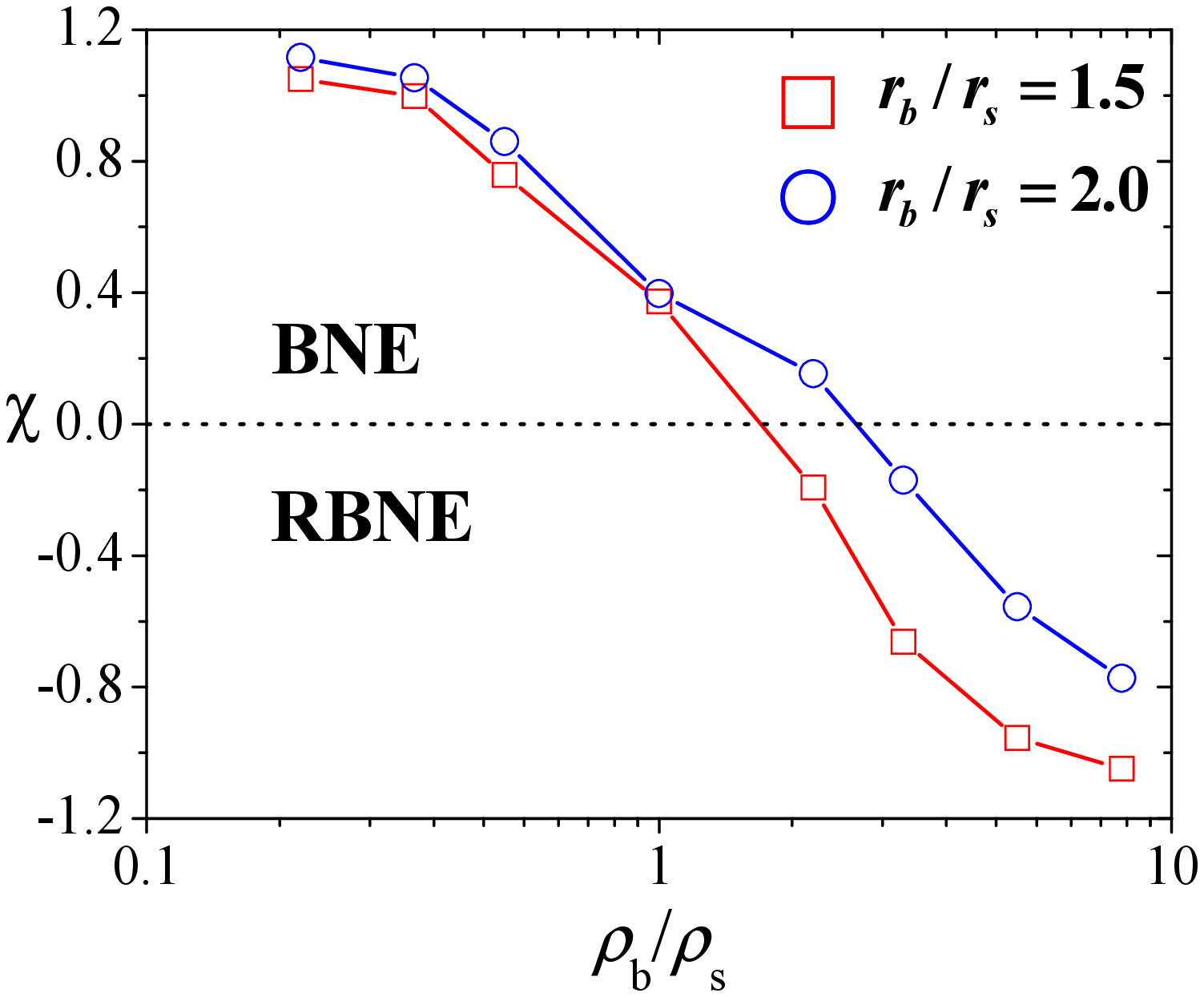}
\caption{Dependence of the segregation index, $\chi$, on the radius ratio of $\rho_b / \rho_s$. The radiuses ratio of $r_b / r_s$ are values being 1.5 (square symbols) and 2.0 (circle symbols), respectively. As $\rho_b / \rho_s$ increases, the system moves from a BNE to a RBNE segregation pattern. The dashed line corresponds to the boundary of BNE and RBNE. The solid line is guided for eyes.}
\label{fig:Figure5}
\end{figure}

To give a quantitative description of the above ring-like spin segregation pattern, we first plot the dependence of the segregation index on the density ratio of $\rho_b / \rho_s$ for different values of radius ratio $r_b / r_s$. In Fig.\ref {fig:Figure5}, at a given radius ratio $r_b / r_s$, the system changes from the S+D-type to the S-D-type when the density ratio of $\rho_b / \rho_s$ grows. The segregation pattern of binary mixtures moves from the BNE to the RBNE. Furthermore, as shown in Fig.\ref {fig:Figure5}, the whole curve shifts upwards as the radius ratio $r_b / r_s$ increases, which means that the switch of the BNE to RBNE needs a higher value of density ratio of $\rho_b / \rho_s$.

\begin{figure}[htbp]
\centering
\vspace {-1.0mm}
\includegraphics[width=7.0cm]{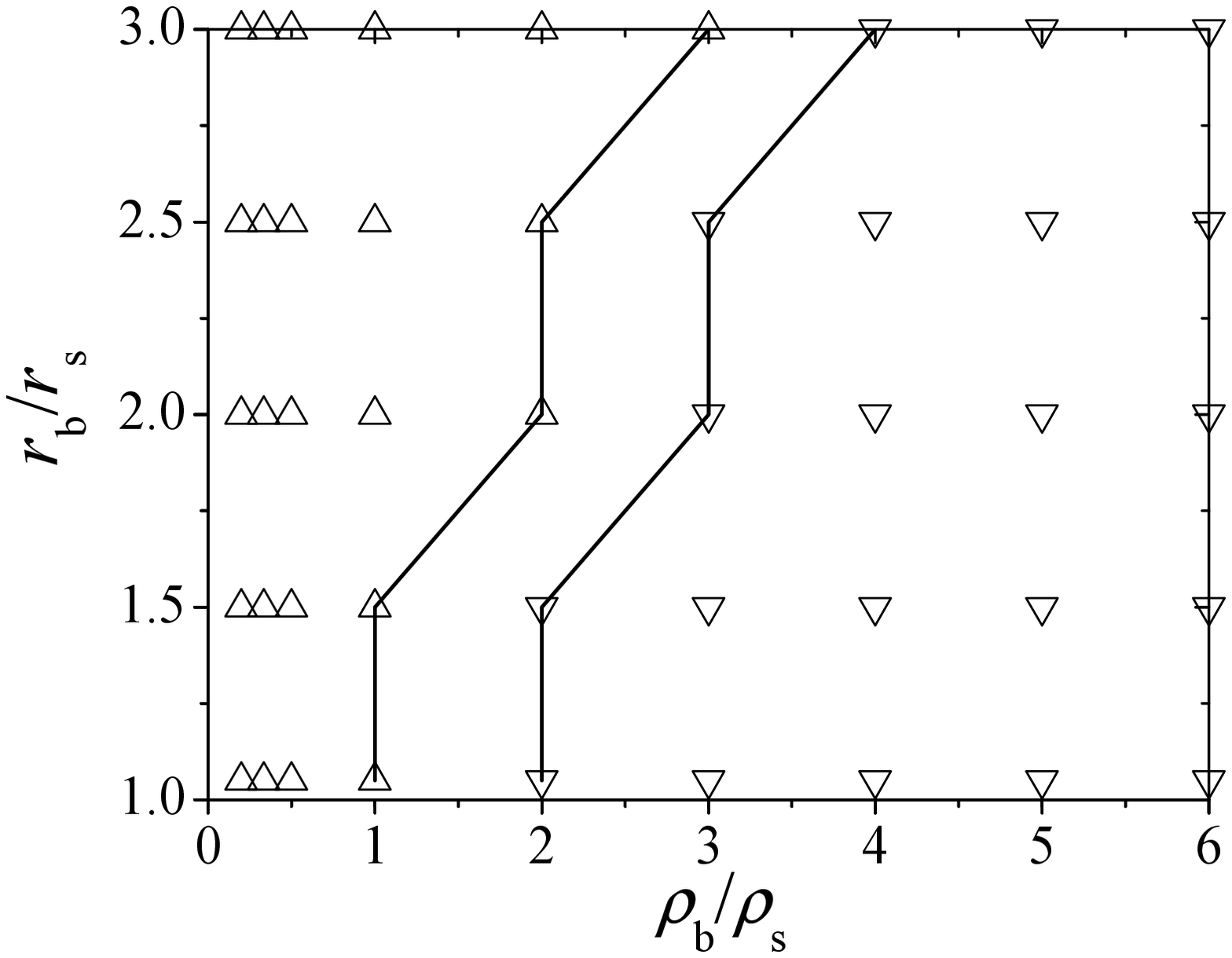}
\caption{Phase diagram of the binary mixtures in the $\rho_b / \rho_s$ - $r_b / r_s$ space. The upward-pointing triangles and downward-pointing triangles represent the segregation patterns being the Brazil-nut effect and the Reverse Brazil-nut effect. The solid lines show the boundaries of two segregation patterns. }
\label{fig:Figure6}
\end{figure}

Fig.\ref {fig:Figure6} summarizes the above findings in the phase diagram of ${\rho_b}/{\rho_s}$-${r_b}/{r_s}$ space by using the segregation index. As shown in Fig.\ref {fig:Figure6}, the BNE and the RBNE are indicated by the symbols of the upward-pointing triangles and downward-pointing triangles, respectively. The solid lines give the boundaries of the BNE and the RBNE, where the absolute values of $\chi$ are actually very small and the corresponding states are almost mixed. In Fig.\ref {fig:Figure6}, it can be easily seen that the BNE appears at the upper left region and the RBNE occurs at the lower right region. Furthermore, the boundary shifts right when the density ratio of bigger particles to smaller ones increases. This result suggests that the BNE is favored for the lower $\rho_b / \rho_s$ and the larger $r_b/r_s$, while the RBNE for the higher $\rho_b / \rho_s$ and the smaller $r_b/r_s$.

The above segregation phenomena due to size and density might seem to result from the two effects, i.e, percolation effect and buoyancy effect. At the case of the centrifugal flow regime, the centrifugal force is always in the radial direction and the outermost region, i.e., the drum wall, is the place of bottom region of granular flowing layer. For an S-type system, the smaller particles are easier to be captured by the voids between the particles in the flowing process and migrate to the bottom region. On the other hand, for a D-type system, the centrifugal force is linear to the mass of particle, i.e., $F_{c} \propto m_p$. It means that the particle of higher density possesses a larger centrifugal force compared to that of lower density. To minimize this centrifugal potential of system, the species of particles with higher density are more likely pushed to the drum wall, i.e., the bottom region. This kind of segregation effect due to the difference of density is here analogous to the buoyancy effect happened in the vertical vibrated binary mixtures. According to the discussions of percolation effect and buoyancy effect, it is reasonable to see what happened in the S+D-type system and the S-D-type system. For an S+D-type system, the percolation effect and buoyancy effect act in the same direction and this combined effects result in a more segregated state. In reverse, the percolation effect and buoyancy effect will compete with each other in an S-D-type system. Thus the final state of binary mixtures might be segregated or mixed, which depends on the combined actions of the percolation effect and buoyancy effect. For example, when the percolation effect overcomes the buoyancy effect, the BNE happens, while when the latter dominates, the RBNE occurs. If the two effects are almost equally, a mixed state will finally yield.

\section{Conclusions}

In this paper, a ring-like spin segregation pattern of binary mixtures in a high-velocity rotational drum is investigated by computer simulations. The radial segregation we observed here is analogous to that a vertical vibrated container, i.e., the Brazil-nut effect, in which the larger particles accumulate towards the center of drum and the smaller ones towards the drum wall, and the Reverse Brazil-nut effect, in which the larger particles migrate towards the drum wall and the smaller ones towards the center of drum. The dependence of the segregation index on the density ratio of $\rho_b / \rho_s$ further quantify the switch between the BNE and The RBNE. The phase diagram in the ${\rho_b}/{\rho_s}$ - ${r_b}/{r_s}$ space shows that the BNE is favored for lower density ratio and larger size ratio of bigger particles to smaller ones and the RBNE for higher density ratio and smaller size ratio of bigger particles to smaller ones. We also propose a simple analysis that the simulated phenomena result from the combined actions of percolation effect and buoyancy effect. When both effects act in same direction, the BNE happens, while when the density ratio of bigger particles to smaller ones is increased high enough and the buoyancy effect exceeds the percolation effect, the RBNE occurs. Since the segregation is of very importance in processing granular mixtures, the results in this paper have potential industrial and theoretical values.

\begin{acknowledgments}
This work is financially supported by the Young Scientists Fund of the National Natural Science Foundation of China (Grant No. 10904070), the National Natural Science Foundation of China (Grant Nos. 10875166 and 10847146) and the NJUST Young Scholar Research Fund (Grant No. 200705).
\end{acknowledgments}

\end{document}